\documentclass[lettersize,journal]{IEEEtran}
\usepackage{amsmath,amsfonts}
\usepackage{algorithmic}
\usepackage{algorithm}
\usepackage{array}
\usepackage[caption=false,font=normalsize,labelfont=sf,textfont=sf]{subfig}
\usepackage{textcomp}
\usepackage{stfloats}
\usepackage{url}
\usepackage{verbatim}
\usepackage{graphicx}
\usepackage{cite}
\usepackage{amssymb}
\usepackage{xcolor}
\usepackage{epsfig}
\usepackage{epstopdf} 
\usepackage{multirow}
\begin{document}

\title{Sequential Task Assignment and Resource Allocation in V2X-Enabled Mobile Edge Computing}

\author{Yufei Ye,~\IEEEmembership{Graduate Student Member,~IEEE}, Shijian Gao,~\IEEEmembership{Member,~IEEE}, Xinhu Zheng,~\IEEEmembership{Member,~IEEE}, and Liuqing Yang,~\IEEEmembership{Fellow,~IEEE}
}



\maketitle

\begin{abstract}
Nowadays, the convergence of mobile edge computing (MEC) and vehicular networks has emerged as a vital enabler for the ever-increasing intelligent onboard applications. This paper proposes a multi-tier task offloading mechanism for MEC-enabled vehicular networks leveraging vehicle-to-everything (V2X) communications. The study focuses on applications with sequential subtasks and explores the collaboration of two tiers. In the Vehicle Tier, the requesting vehicle (RV)-service vehicle (SV) matching scheme and the inter-vehicle collaborative computation are studied, with joint optimization of task offloading decision, communication, and computing resource allocation to minimize energy consumption while satisfying delay requirements. In the Roadside Unit (RSU) Tier, collaboration among RSUs is investigated to further address multi-access issues of uplink subchannels and computing resources for serving unmatched RVs. To tackle this intricate problem, a layered optimization framework is first proposed to obtain task offloading decisions and optimal continuous resource allocation, after which a subchannel allocation scheme is designed to recover the discrete solution with low complexity. Extensive experiments are conducted to demonstrate that the proposed method reduces average energy consumption by at least 15\% compared with recent utility maximization and energy cost minimization benchmarks under varying task delay requirements and vehicle scales.
\end{abstract}



\section{Introduction}

In recent years, rapid advancements in intelligent vehicles, vehicular networks, and autonomous driving have significantly enhanced driving safety, traffic efficiency, and comfort for passengers \cite{xcheng2022mmwave, xzheng2022multivehicle}. Equipped with advanced sensors, such as cameras, LiDARs, and radars, onboard devices can support various intelligent applications by processing diverse types of collected data, such as object detection, multi-modal perception data fusion \cite{droy2023multi}, and tracking \cite{xcheng20233d}.

Vision-based onboard applications powered by image/video processing and deep neural networks (DNNs) are typically computation-intensive and delay-sensitive to ensure timeliness and security. The resource-limited onboard devices often struggle to satisfy these stringent requirements. Fortunately, mobile edge computing (MEC) \cite{ymao2017asurvey} has emerged as a promising paradigm by providing efficient computing services in the vicinity of vehicles. Recent advances in MEC \cite{wfan2024collaborative,jyan2024bayesian} have stimulated research on task offloading and resource allocation strategies tailored to intelligent vehicular applications \cite{xcheng2022integrated}. MEC servers deployed on roadside units (RSUs) provide substantial computing capacity via vehicle-to-infrastructure (V2I) links \cite{schen2025an} while inevitably confront severe multi-access resource competition. Complementarily, rapid vehicle-to-vehicle (V2V) links also enable the resource-sharing and collaborative computation among vehicles \cite{qwu2024delay}. Powering these resource-intensive applications incurs significant energy consumption. Consequently, a sophisticated task and resource scheduling scheme is imperative to foster energy-efficient, eco-friendly computing services while ensuring timely task completion.

Vehicular task offloading commonly adopts either binary or partial offloading modes. Binary offloading models each application task as an indivisible entity executed entirely by an onboard device or an MEC server. A multi-objective problem of vehicular task offloading is identified and addressed by a hybrid genetic-based algorithm in \cite{schen2025an}, while \cite{xliu2025large} studies the dimensionality reduction of large-scale multi-objective problem. In \cite{yzhang2025latency}, a vehicle-RSU cooperative framework is proposed to minimize task delay of autonomous vehicles. A multi-hop offloading strategy is studied in \cite{whuang2025mmto} to maximize system profit via an incentive scheme. The offloading success rate is enhanced via a vehicle trajectory prediction scheme in \cite{hguo2022v2v}. Based on the alternating direction method of multipliers, \cite{xwang2024amtos} and \cite{wshu2024an} investigate cloud-edge-end collaboration to provide heterogeneous services for vehicles. The energy cost of automated guided vehicle patrol is minimized through intra-cluster information sharing in \cite{ywu2024anovel}. However, assigning an entire task to a single device often leads to workload imbalance and excessive energy consumption on overloaded devices, while failing to fully utilize the distributed computing resources across the system.

In contrast, partial offloading treats tasks as divisible and distributable among multiple devices or servers. A strategy leveraging nearby vehicles to assist platoons is proposed in \cite{qwu2024delay}, while \cite{hzhang2024partial} designs a graph coloring algorithm to address channel allocation in a multi-user V2X network. \cite{stian2025partial} proposes a deep deterministic policy gradient framework to optimize task splitting and resource allocation. A multi-tier uncrewed aerial vehicle (UAV) edge computing system is developed in \cite{yye2025multitier} to minimize task execution delay while stabilizing UAV energy reserves. An adaptive offloading method is proposed in \cite{sli2024road} to handle uncertain channel states and network topology, while \cite{nfofana2025intelligent} achieves balanced optimization of task delay, expense, and success rate under fluctuating network states. Although partial offloading enhances computational efficiency, it largely overlooks the interdependencies among subtasks. Moreover, certain subtasks, such as individual layers in DNNs, are inherently indivisible. Consequently, the practicality of dividing an application task into arbitrary continuous proportions remains limited.

Some studies consider subtask dependencies within each application. In \cite{lzhao2025dual}, the application success rate, delay, energy cost, and caching hit rate are jointly optimized considering both task and service dependencies. A graph neural network–based algorithm is proposed in \cite{jwu2025dependency} to adapt to dynamic vehicle and server scales, while \cite{lzhao2024stackelberg} maximizes the utilities of both vehicles and servers via Stackelberg game theory. \cite{hliu2024towards} develops a distributed task offloading strategy with partial agent observation, while \cite{lzhao2024meson} designs a vehicle mobility detection scheme to facilitate stable V2I links. An adaptive task offloading and channel occupation strategy based on meta reinforcement learning is proposed in \cite{pdai2024meta}. The authors in \cite{zwang2024low} develop a low-complexity greedy offloading method to reduce task processing delay, while the task offloading efficiency integrating task delay and energy cost is maximized in \cite{qshen2022dependency}. However, these works largely neglect the multi-access challenges of discrete V2I subchannels and computing resources among servers, which are critical when servers simultaneously assist multiple vehicles. Furthermore, most existing schemes dispatch dependent subtasks within an application across multiple devices in a scattered manner. This necessitates frequent wireless transmissions, thereby increasing communication latency and the risk of encountering poor channel conditions. Consequently, the available computation time is significantly shortened, forcing devices to operate at higher processor frequencies and consume excessive energy to satisfy strict deadlines.

\begin{table}[!t]
\caption{Comparison with Related Studies}
\label{tab:comparison}
\centering
\begin{tabular}{|m{1.2cm}<{\centering}|m{1.7cm}<{\centering}|m{1.9cm}<{\centering}|m{1.8cm}<{\centering}|}
\hline
\multirow{2}{*}[-1.7em]{\textbf{Reference}} & \multicolumn{3}{c|}{\textbf{Research Elements}} \\
\cline{2-4}
 & \textbf{Subtask Dependency} & \textbf{Multi-Vehicle Computing Resource Coordination} & \textbf{V2I Multi-Vehicle Access} \\
\hline
\cite{schen2025an} & $\times$ & $\times$ & $\times$ \\
\hline
\cite{qwu2024delay, xliu2025large, yzhang2025latency} & $\times$ & \checkmark & $\times$ \\
\hline
\cite{whuang2025mmto, hguo2022v2v} & $\times$ & $\times$ & $\times$ \\
\hline
\cite{xwang2024amtos, wshu2024an, ywu2024anovel, hzhang2024partial, stian2025partial, yye2025multitier} & $\times$ & \checkmark & $\times$ \\
\hline
\cite{sli2024road} & $\times$ & $\times$ & $\times$ \\
\hline
\cite{nfofana2025intelligent} & $\times$ & \checkmark & $\times$ \\
\hline
\cite{lzhao2025dual, jwu2025dependency} & \checkmark & $\times$ & $\times$ \\
\hline
\cite{lzhao2024stackelberg} & \checkmark & \checkmark & $\times$\\
\hline
\cite{hliu2024towards, lzhao2024meson, pdai2024meta, zwang2024low, qshen2022dependency} & \checkmark & $\times$ & $\times$ \\
\hline
Our work & \checkmark & \checkmark & \checkmark \\
\hline

\end{tabular}
\end{table}



In this paper, we propose a multi-tier task offloading mechanism facilitated by vehicle-to-everything (V2X) communications \cite{mhcgarcia2021atutorial} for onboard applications with sequential subtasks in an MEC-enabled vehicular network. The computer vision-based vehicular applications (e.g., vehicle/pedestrian detection, traffic sign recognition, environment perception) inherently exhibit sequential execution dependencies among their subtasks, where each typically corresponds to an operation in processing pipeline or a DNN layer \cite{ymao2017asurvey}. In the Vehicle Tier, to fully exploit rapid V2V communications and alleviate computational burdens of RSUs, we design a requesting vehicle (RV)-service vehicle (SV) matching scheme that aims at finding qualified SVs to assist as many RVs as possible in computing tasks that they cannot complete on time with local computing capacity. By jointly optimizing task offloading decisions and allocation strategies for communication and computing resources, we achieve the goal of minimizing vehicle energy consumption while satisfying delay requirement of each application task. The RVs that fail to be served in the Vehicle Tier will offload their tasks to RSUs for collaborative computation, which will be investigated in the RSU Tier. Beyond the optimization aspects of the Vehicle Tier, we further explore the multi-access issues of V2I subchannels and computing resources for RSUs serving multiple RVs, which is formulated as an intricate problem involving multifaceted discrete and continuous optimization variables. To tackle this problem, we first propose a layered optimization framework to derive the task offloading decisions alongside optimal continuous solutions of V2I transmission delay, V2I bandwidth allocation, and RSU computing resource allocation via continuous relaxation. Afterwards, a low-complexity V2I subchannel allocation scheme is designed to recover the near-optimal discrete solution. Additionally, each application task involves only one wireless transmission of intermediate data to limit communication delay and reduce sensitivity to poor channel conditions. To the best of our knowledge, this is the first work that simultaneously addresses subtask dependencies, demand-aware multi-vehicle computing resource coordination, and realistic discrete V2I subchannel allocation, proposing a holistic solution for these critical issues. We summarize the comparison with related works in TABLE~\ref{tab:comparison}. The main contributions of this work are outlined as follows:

\begin{enumerate}
    \item A novel multi-tier task offloading mechanism is proposed for onboard applications with sequential subtasks. In the Vehicle Tier, an efficient RV-SV matching scheme is devised to maximize the number of matched RVs. Then we jointly optimize task offloading decisions, transmission delay, and computing resource allocation to minimize vehicle energy consumption while meeting delay requirements.
    
    \item Beyond the optimization dimensions in the Vehicle Tier, we further address multi-access challenges for unmatched RVs in the RSU Tier. We propose a layered framework followed by a subchannel allocation scheme to jointly determine task offloading decisions, V2I subchannel allocation and transmission delays, and RSU computing resource allocation.
    
    \item Extensive experiments demonstrate that the proposed method reduces average energy consumption by at least 15\% compared with benchmarks under varying delay constraints and vehicle scales. Additionally, we comprehensively analyze the impact of diverse system parameters on overall performance.
    
\end{enumerate}

The remainder of this article is organized as follows. We present the system model in Section II. The problem formulation, solving process, and complexity analysis for the methods of the vehicle and RSU tiers are detailed in Section III and Section IV, respectively. The experiment evaluations and analysis are elaborated in Section V. We conclude this paper in Section VI.


\section{System Model}

In this section, we introduce the proposed system model for two-tier edge computing, which consists of the network model, application task model, communication model, and computation model.

\subsection{Network Model}

\begin{figure}[!t]
\centering
\includegraphics[width=3.4 in]{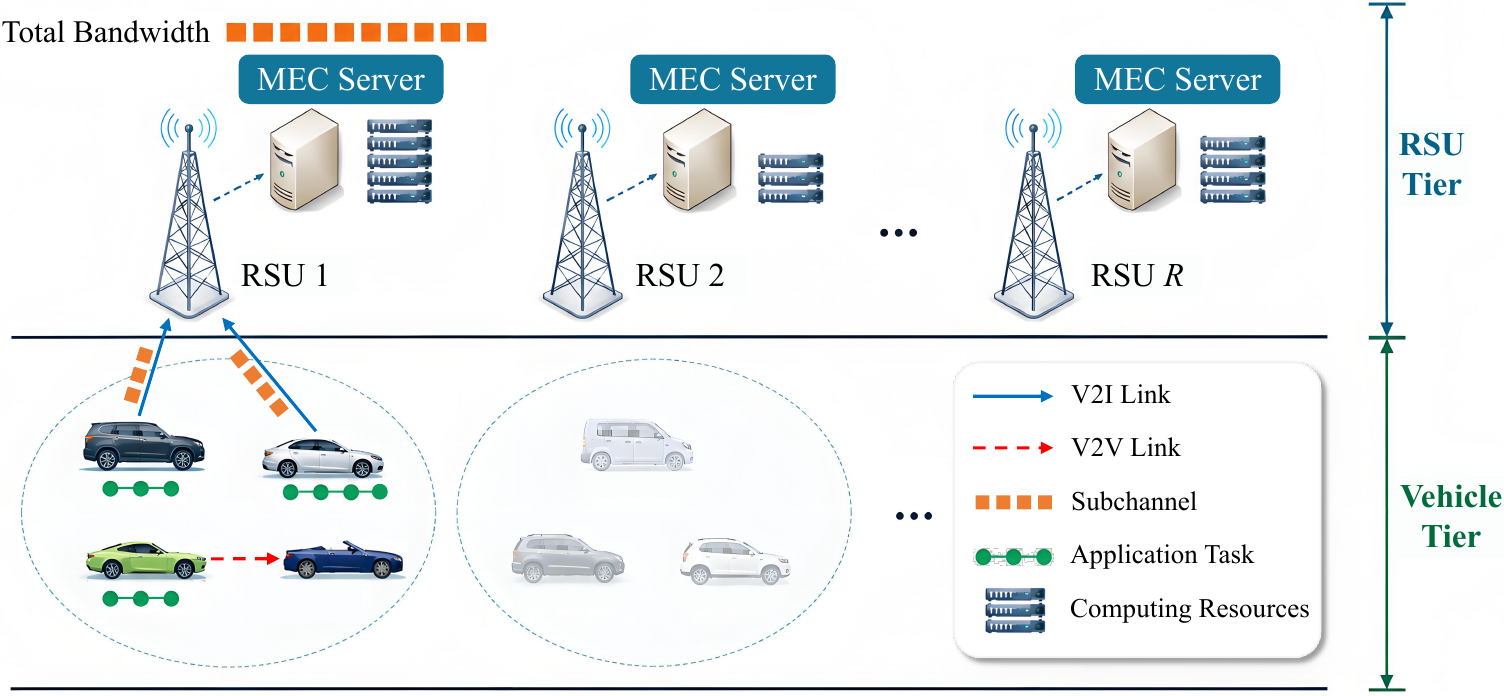}
\caption{Illustration of the two-tier MEC-enabled vehicular network.}
\label{system_model}
\end{figure}

The proposed system architecture is illustrated in Fig.~\ref{system_model}. It encompasses the Vehicle Tier and the RSU Tier in MEC-enabled vehicular network. RSUs equipped with MEC servers are connected via wired links, where different available computing resources are represented by varying numbers of chips. We signify RSU set by $\mathcal{R}=\{1, ..., r, ...,R\}$. Within the service coverage of current RSU$_1$, we consider two types of vehicles: (1) requesting vehicles (RVs) that generate tasks but are unable to complete their tasks on time due to limited onboard computing capabilities, denoted by $\mathcal{N}=\{1, ..., n, ...,N\}$, and (2) idle vehicles (IVs) with surplus computing resources, denoted by $\mathcal{IV}=\{1,2, ..., i, ...,I\}$. A central controller is deployed on the current RSU$_1$, which collects the
system state informations (e.g., vehicle locations and velocities, task profiles, computing capabilities of vehicles and servers), executes the optimization algorithms, and disseminates the task and resource scheduling decisions to the corresponding devices. We focus on application tasks composed of sequential subtasks in this work, as shown by green circle strings. The Cartesian coordinate is adopted, with the positive $x$-axis aligned with vehicle driving direction and the positive $y$-axis perpendicular to the road. We denote the horizontal positions of RV$_n$, IV$_i$, and RSU$_r$ by $\boldsymbol{L_n} = (x_n, y_n)$, $\boldsymbol{L_i} = (x_i, y_i)$, and $\boldsymbol{L_r} = (x_r, y_r)$, respectively. The height of RSU is $H_r$. The velocities of RV$_n$ and IV$_i$ are signified as $v_n$ and $v_i$. The total bandwidth $B$ of RSU$_1$ is divided into $b$ subchannels for V2I communications.

We first match as many RVs as possible with suitable IVs in the Vehicle Tier to assist in task computation, leveraging fast V2V communications and alleviating computational burdens of RSUs. This will be elaborated in Section III. The IVs that are ultimately selected to assist RVs will be referred to as the service vehicles (SVs), denoted by set $\mathcal{S}=\{1, ..., s, ...,S\}$, and we have $\mathcal{S} \subseteq \mathcal{IV}$. Evidently, a vehicle cannot act as both an RV and an SV at the same time, but it may switch between these two roles in different temporal context. For the RVs that fail to be successfully matched in the Vehicle Tier, they will offload their application tasks to RSUs, which will then collaboratively compute tasks for RVs. For each application task, RSUs process its subtasks sequentially along the vehicle driving direction. These will be investigated in Section IV. The applications considered in this study are object detection and recognition tasks in traffic environments, such as vehicle/pedestrian/obstacle detection and traffic sign recognition, etc. The results are typically the bounding box coordinates or the object categories, which can be represented by a small amount of data \cite{droy2023multi}. Therefore, similar to \cite{schen2025an, qwu2024delay}, the size of result data is neglected. For clarity of presentation, the main notations and corresponding definitions in this research are summarized in TABLE~\ref{tab:notations}.

\begin{table}[!t]
\caption{Main Notations and Definitions}
\label{tab:notations}
\centering
\begin{tabular}{|m{0.8cm}<{\centering}|m{1.0cm}<{\centering}|m{5.5cm}<{\centering}|}
\hline
Related Tier & Notations & Definitions\\
\hline
\multirow{8}{*}{General} & $C^n_m$ & Computation workloads of the $m\mbox{-}th$ subtask in the application task of RV$_n$\\
\cline{2-3} &
$W^n_{m-1,m}$ & Amount of intermediate data between subtask $m-1$ and $m$ in the application task of RV$_n$\\
\cline{2-3} &
$T_{n,max}$ & Tolerable delay of the application task generated by RV$_n$\\
\cline{2-3} &
$W_0(x)$ & The principal branch of the Lambert W function \cite{rmcorless1996on} satisfying $W_0(x)e^{W_0(x)}=x$ and $W_0(x)\geq -1$\\
\hline
\multirow{14}{*}{Tier-1} & $m_s^n$ & The first subtask of RV$_n$ computed by SV$_s$ \\
\cline{2-3} &
$B_{V2V}$ & Bandwidth for V2V communication\\
\cline{2-3} &
$f_m^n$ & Computing resources allocated by RV$_n$ to its $m\mbox{-}th$ subtask\\
\cline{2-3} &
$f_m^s$ & Computing resources allocated by SV$_s$ to the $m\mbox{-}th$ subtask of RV$_n$ it serves\\
\cline{2-3} &
$F_n$ & Maximal computing resources of RV$_n$\\
\cline{2-3} &
$F_s$ & Maximal computing resources of SV$_s$\\
\cline{2-3} &
$\tau_{n,s}$ & Data transmission delay from RV$_n$ to SV$_s$\\
\cline{2-3} &
$\tau_{n,m}^c$ & Computation delay of $m\mbox{-}th$ subtask of RV$_n$\\
\cline{2-3} &
$E_{n,s}^t$ & Data transmission energy consumption from RV$_n$ to SV$_s$\\
\cline{2-3} &
$E_{n,m}^c$ & Computation energy consumption of the $m\mbox{-}th$ subtask of RV$_n$\\
\hline
\multirow{18}{*}{Tier-2} & $m_r^n$ & The first subtask of RV$_n$ computed by RSU$_r$\\
\cline{2-3} &
$B$ & Total bandwidth of the current RSU\\
\cline{2-3} &
$B_0$ & Bandwidth of each subchannel\\
\cline{2-3} &
$b_n$ & Number of subchannels allocated to the V2I link between RV$_n$ and the current RSU\\
\cline{2-3} &
$f^r_{n,m}$ & Computing resources allocated by RSU$_r$ to the $m\mbox{-}th$ subtask of RV$_n$\\
\cline{2-3} &
$F_r$ & Maximal computing resources of RSU$_r$\\
\cline{2-3} &
$\tau_n$ & Data transmission delay from RV$_n$ to the current RSU\\
\cline{2-3} &
$\tau_{r,n}$ & Data transmission delay of RV$_n$ from RSU$_r$ to RSU$_{r+1}$\\
\cline{2-3} &
$\tau^c_{n,m}$ & Computation delay of $m\mbox{-}th$ subtask of RV$_n$\\
\cline{2-3} &
$E_n^t$ & Data transmission energy consumption from RV$_n$ to the current RSU\\
\cline{2-3} &
$E^t_{r,n}$ & Data transmission energy consumption of RV$_n$ from RSU$_r$ to RSU$_{r+1}$\\
\cline{2-3} &
$E_{n,m}^c$ & Computation energy consumption of the $m\mbox{-}th$ subtask of RV$_n$\\
\hline

\end{tabular}
\end{table}


\subsection{Application Task Model}

\begin{figure}[!t]
\centering
\includegraphics[width=3.3 in]{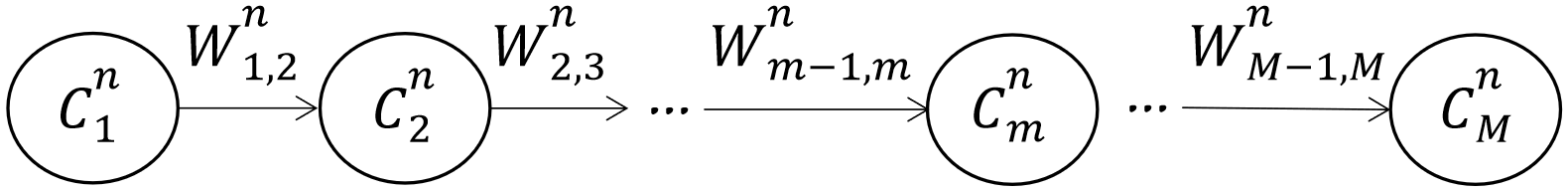}
\caption{Application task model.}
\label{task_model}
\end{figure}

We target at application tasks whose subtasks exhibit sequential dependencies. Only after the previous subtask is completed and its output data is obtained can the next subtask begin to be computed. Thus, we model each application task as a unidirectional graph as shown in Fig.~\ref{task_model}. Each node represents a subtask, and each edge shows the dependency between adjacent subtasks. We use set $\mathcal{M}^n=\{1, ..., m, ...,M^n\}$ to denote the subtasks in the application task of RV$_n$, with the total tolerable delay $T_{n,max}$. The computational workloads of the $m\mbox{-}th$ subtask of RV$_n$ is $C^n_m$ CPU cycles. The size of intermediate data between the $m-1\mbox{-}th$ and $m\mbox{-}th$ subtask is $W^n_{m-1,m}$ bits.


\subsection{Communication Model}

In the Vehicle Tier, for each RV$_n$, let $1, ..., m_s^n-1$ denote the indices of its subtasks computed locally on this RV, and the $m_s^n\mbox{-}th$ to $M^n\mbox{-}th$ subtasks are computed on its matched SV$_s$. In the RSU Tier, let $m_r^n , ..., m_{r+1}^n-1$ represent the indices of RV$_n$'s subtasks computed on RSU$_r$. If RSU$_r$ is the last RSU that processes RV$_n$'s application task, it computes the $m_r^n\mbox{-}th$ to $M^n\mbox{-}th$ subtasks.

Based on the above symbolization, for the Vehicle Tier, RV$_n$ offloads subtasks starting from $m_s^n$. Therefore, it should transmit the input data of the $m_s^n\mbox{-}th$ subtask with the size of $W_{m_s-1,m_s}^n$ to its matched SV$_s$. Similar to \cite{xwang2024amtos,nfofana2025intelligent}, we adopt Orthogonal Frequency Division Multiplexing (OFDM) and Orthogonal Frequency Division Multiple Access (OFDMA) technologies for V2V and V2I communications, respectively. Each communication link occupies mutually orthogonal subbands, which exhibits strong resistance to signal interference. We signify the bandwidth for V2V communication as $B_{V2V}$, the V2V data transmission delay from RV$_n$ to SV$_s$ as $\tau_{n,s}$, the corresponding transmission energy consumption as $E^t_{n,s}$, and the power spectrum density of noise at SV as $N_0$. Therefore, based on Shannon capacity formula, we can obtain
\begin{equation*}
W_{m_s-1,m_s}^n = \tau_{n,s}B_{V2V}\ln\left(1+\dfrac{E_{n,s}^t 10^{-\frac{\phi_{n,s}}{10}} h_{n,s}}{\tau_{n,s}B_{V2V}N_0}\right), \tag{1}
\end{equation*}
where $\phi_{n,s}=63.3+17.7\lg(d_{n,s})$ \cite{hguo2022v2v} is the path loss between RV$_n$ and SV$_s$, $d_{n,s} = \sqrt{(x_n - x_s)^2 + (y_n - y_s)^2}$ denotes the distance between them, and $h_{n,s}$ symbolizes the channel fading factor.

Regarding the RSU Tier, each RV$_n$ needs to transmit the initial task data (i.e., the input data of the first subtask) with the size of $W^n_0$ to RSU$_1$. We denote the V2I data transmission delay from RV$_n$ to RSU$_1$ as $\tau_n$, the corresponding transmission energy as $E_n^t$, the power spectrum density of noise at RSU$_1$ as $N_0$, and the bandwidth allocated to the link between RV$_n$ and RSU$_1$ as $b_nB_0$, where $B_0$ is the bandwidth of each subchannel and $b_n$ is the number of subchannels allocated to this link. Then we have
\begin{equation*}
W^n_0=\tau_n b_nB_0\ln\left(1+\dfrac{E_n^t \phi_n h_n}{\tau_n b_nB_0N_0}\right), \tag{2}
\end{equation*}
where $\phi_n = d_{n}^{-\delta}$ \cite{nfofana2025intelligent} is defined as the path loss, with $d_{n} = \sqrt{(x_n - x_r)^2 + (y_n - y_r)^2 + H_r^2}$ and $\delta$ denoting the distance between them and path loss factor, respectively.

Based on (1) and (2), the expression of V2V and V2I transmission energy consumption $E_{n,s}^t$ and $E_n^t$ can be rewritten as follows respectively
\begin{equation*}
E_{n,s}^t=\frac{B_{V2V}N_0\tau_{n,s}}{10^{-\frac{\phi_{n,s}}{10}} h_{n,s}}\left(e^{\frac{W_{m_s-1,m_s}^n}{\tau_{n,s}B_{V2V}}}-1\right), \tag{3}
\end{equation*}
\begin{equation*}
E_n^t=\dfrac{b_nB_0N_0\tau_n}{\phi_n h_n}\left(e^{\frac{W^n_0}{\tau_n b_nB_0}}-1\right). \tag{4}
\end{equation*}

We model the wired transmission among RSUs as follows. Suppose that the required energy for each RSU to transmit one bit of data to the next RSU through wired link is $E_0$ and the delay for transmitting one bit is $\tau_0$. Then the energy consumption for RSU$_r$ to transmit the intermediate data of RV$_n$ to RSU$_{r+1}$ can be expressed by $E^t_{r,n}=E_0 W^n_{m_{r+1}-1,m_{r+1}}$ and the corresponding data transmission delay is $\tau_{r,n}=\tau_0W^n_{m_{r+1}-1,m_{r+1}}$.

\emph{Remark:} The realistic vehicular networks exhibit inherent complexity and uncertainty. To clearly highlight the core research problem of energy-efficient task and resource scheduling, following most existing works, the detailed channel state fluctuations and signal processing procedures in the physical layer are reasonably abstracted.


\subsection{Computation Model}

Without loss of generality, we adopt general CPU computation model. For the Vehicle Tier, we denote the computing resource (i.e., CPU frequency) allocated by each pair of RV$_n$ and SV$_s$ to process the $m\mbox{-}th$ subtask of this RV$_n$ as $f_m^n$ and $f_m^s$, respectively. The computation delay is given by
\begin{equation*}
\tau_{n,m}^c=\frac{C_m^n}{f_m^n}, \hspace{0.1cm} 1\leq m \leq m_s^n-1, \tag{5a}
\end{equation*}
\begin{equation*}
\tau_{n,m}^c=\frac{C_m^n}{f_m^s}, \hspace{0.1cm} m_s^n\leq m \leq M^n. \tag{5b}
\end{equation*}
The computation energy consumption of the $m\mbox{-}th$ subtask for RV$_n$ can be expressed as
\begin{equation*}
E_{n,m}^c=\kappa_n C_m^n\left(f_m^n\right)^2, \hspace{0.1cm} 1\leq m \leq m_s^n-1, \tag{6a}
\end{equation*}
\begin{equation*}
E_{n,m}^c=\kappa_s C_m^n\left(f_m^s\right)^2, \hspace{0.1cm} m_s^n\leq m \leq M^n, \tag{6b}
\end{equation*}
where $\kappa_n$ and $\kappa_s$ signify the energy efficiency coefficient \cite{ymao2017asurvey} of RV$_n$'s and its SV$_s$'s processor, respectively, depending on the chip structure. In the RSU Tier, we denote the computing resource allocated by RSU$_r$ to the $m\mbox{-}th$ subtask of RV$_n$ as $f^r_{n,m}$ if $m_r^n\leq m \leq m_{r+1}^n-1$. The computation delay and energy consumption are respectively given by
\begin{equation*}
\tau^c_{n,m}=\frac{C_m^n}{f^r_{n,m}}, \tag{7a}
\end{equation*}
\begin{equation*}
E_{n,m}^c=\kappa_r C_m^n\left(f^r_{n,m}\right)^2, \tag{7b}
\end{equation*}
where $\kappa_r$ is the energy efficiency coefficient of RSU$_r$.


\section{Tier-1: Collaborative Computation Between RV and SV in the Vehicle Tier}

In this section, we elaborate on the investigation of the Vehicle Tier. We first present the RV-SV matching algorithm, followed by the problem formulation of the Vehicle Tier. Subsequently, we detail the solution to this problem and conduct complexity analysis for our solution.

\subsection{RV-SV Matching}

\begin{algorithm}[!t]
    \caption{RV-SV Matching Algorithm.}
    \label{alg:matching}
    \renewcommand{\algorithmicrequire}{\textbf{Input:}}
    \renewcommand{\algorithmicensure}{\textbf{Output:}}
    
    \begin{algorithmic}[1]
        \REQUIRE $\mathcal{N}$, $\mathcal{IV}$, $\mathcal{M}^n$, $\boldsymbol{L_n}$, $\boldsymbol{L_i}$, $v_n$, $v_i$, $D_{V2V}$, $C_m^n$, $W_0^n$, $F_i$.   
        \ENSURE RV-SV matching result $\boldsymbol{A}^*$.     

        \STATE Initialize the adjacency matrix $\boldsymbol{A}$ for RVs and IVs. Initialize the adjacency indicator $a_{n,i}$ for each pair of RV$_n$ and IV$_i$ to 0.

        \FOR{$n \in \mathcal{N}$}
        
            \FOR{$i \in \mathcal{IV}$}

                \STATE Calculate distance $d_{n,i}$ between the two vehicles.

                \STATE Calculate task computation delay with $T_{n,i} = \sum_{m \in \mathcal{M}^n} C_m^n / F_i$.
            
                \IF{$d_{n,i} < D_{V2V}$ and $T_{n,i} < T_{n,max}$}
                
                    \STATE $a_{n,i} \leftarrow 1$.

                \ELSIF{$d_{n,i} == D_{V2V}$}

                    \IF{$x_i > x_n, v_i < v_n, T_{n,i} < T_{n,max}$ or $x_n > x_i, v_n < v_i, T_{n,i} < T_{n,max}$}

                        \STATE $a_{n,i} \leftarrow 1$.

                    \ENDIF
                
                \ENDIF

            \ENDFOR

        \ENDFOR

        \STATE Execute the Hungarian algorithm based on the current adjacency matrix to obtain the final maximum matching.
        
        \RETURN RV-SV matching result.

    \end{algorithmic}
\end{algorithm}

In this subsection, we introduce the RV-SV matching algorithm, which aims to match as many RVs as possible with qualified IVs within the system consisting of $N$ RVs and $I$ IVs. After the matching process, the IVs that are selected to serve as helpers for RVs will become SVs.

We jointly consider computing capabilities, positions, and velocities of RVs and IVs, the data size, computational workloads, and delay requirements of RVs' application tasks for link stability and on-time task computation \cite{yfan2023radar}. The maximum V2V communication range is denoted by $D_{V2V}$. The matching algorithm consists of the following two parts:

(1) \textbf{Candidate RV-IV Matching Identification:} Only the RV-IV pair that simultaneously satisfies the following two conditions can be considered as a candidate match:
\begin{itemize} 

\item \textit{V2V Communication Range Condition}: The distance between the two vehicles is less than $D_{V2V}$, or the distance is equal to $D_{V2V}$ but the vehicle positioned behind is driving at a higher speed than the one in front, implying that the two vehicles are approaching each other. Note that, to prevent communication interruption, the extreme case where the distance equals $D_{V2V}$ while their speeds are identical is excluded.

\item \textit{IV Computing Resource Condition}: When the IV utilizes its maximum computing resources $F_i$ to process this RV's task, the task computation delay should be shorter than the delay requirement of this task.

\end{itemize}
Through this process, we obtain all potential RV-IV matches.

(2) \textbf{Maximum Matching for RVs:} We employ the Hungarian algorithm \cite{mzalam2022multi} to perform the maximum matching for RVs based on the current set of candidate RV-IV matches. The selected IVs in the final matching are promoted to SVs. The detail of the Hungarian algorithm is omitted due to space limitation. In this manner, we maximize the number of RVs that can find qualified SVs under the conditions of the Vehicle Tier. The steps of matching process are summarized in \textbf{Algorithm~\ref{alg:matching}}. Subsequently, we further optimize the task offloading decision and allocation strategies for communication and computing resources of each RV-SV pair.


\subsection{Problem Formulation of the Vehicle Tier}
After determining the RV-SV matching, in this subsection, we formulate the problem in the Vehicle Tier to jointly optimize V2V task offloading decision, transmission delay, and computation resource allocation. The objective is to minimize the energy consumption of vehicles while ensuring the delay requirements of RVs' application tasks. The optimization problem $\mathcal{P}_1$ of Tier-1 is formulated as

$\mathcal{P}_1:$ \textbf{Collaborative Computation of Vehicles}

\vspace{-0.4cm}
\begin{align*}
\min _{\substack {\mathbf{m_s^n}, \mathbf{\tau_{n,s}},\\ \mathbf{f_m^V}}} &\sum_{s=1}^{S} \left[w_n \hspace{-1mm} \left ( E_{n,s}^t + \sum _{m=1}^{m_s^n-1} E_{n,m}^c \right) + w_s \sum_{m=m_s^n}^{M^n} E_{n,m}^c \right]\\
\text {s.t. } \hspace{0.1cm} &1 \leq m_s^n \leq M^n, \forall s \in \mathcal {S}, \tag{8a}\\
& \sum_{m=1}^{M^n} \tau_{n,m}^c +\tau_{n,s} \leq T_{n,max}, \forall s \in \mathcal {S}, \tag{8b}\\
& \tau_{n,s} \geq 0, \forall s \in \mathcal {S}, \tag{8c}\\
& 0 \leq f_m^n \leq F_n, m=1,...,m_s^n-1, \forall s \in \mathcal {S},\tag{8d}\\
& 0 \leq f_m^s \leq F_s, m=m_s^n,...,M^n, \forall s \in \mathcal {S}. \tag{8e}
\end{align*}

In the objective function, the first two parts are the V2V data transmission energy and the computation energy for the first $m_s^n-1$ subtasks consumed by each RV, respectively, while the third part contains the computation energy of each SV for processing remaining subtasks. $w_n$ and $w_s$ are the weights of RV$_n$ and SV$_s$, respectively. Since not all RVs can be successfully matched in the Vehicle Tier, each RV–SV pair is indexed by the corresponding SV. Regarding the optimization variables, set $\mathbf{m_s^n}=\{m_s^n|s \in \mathcal{S}\}$ encompasses the offloading decisions for RVs, $\mathbf{\tau_{n,s}}=\{\tau_{n,s}| s \in \mathcal{S}\}$ contains the V2V transmission delays, and $\mathbf{f_m^V} = \left\{ f_m^n|m=1,...,m_s^n-1 \right\} \cup \left\{f_m^s|m=m_s^n,...,M^n \right\}$, $\forall s \in \mathcal{S}$ includes the computing resources allocated by the two types of vehicles to each subtask. As for the constraints, (8a) limits the range of each $m_s^n$ and (8b) guarantees the total execution delay of each application task does not exceed its requirement. (8c) ensures the nonnegativity of transmission delay. The computing resources allocated by the two types of vehicles to each subtask should not be greater than their respective capacities are guaranteed by (8d) and (8e).


\subsection{Problem Solving}
To cope with this mixed-integer nonlinear programming (MINLP) problem, we first fix the integer variable $m_s^n$ and optimize the remaining two types of continuous variables. Afterwards, the optimal offloading decision $m_s^{n*}$ for each RV can be determined via the linear search method.

For the notational convenience, the following definitions are introduced for each RV–SV pair:
\begin{align*}
f^{V}_{m} \triangleq \begin{cases} f_m^n, & m=1,\ldots, m_s^n-1, \\ f_m^s, & m=m_s^n,\ldots, M^n, \end{cases}\tag{9}
\end{align*}
\begin{align*}
F_V \triangleq \begin{cases} F_n, & m=1,\ldots, m_s^n-1, \\ F_s, & m=m_s^n,\ldots, M^n. \end{cases}\tag{10}
\end{align*} 
Through the above definitions, we merge equations (8d) and (8e) into
\begin{align*}
0 \leq f^{V}_{m} \leq F_V, m \in \mathcal{M}^n. \tag{11}
\end{align*}

\textbf{Proposition 1:} \textit{Problem $\mathcal{P}_1$ is a convex optimization problem with fixed task offloading decisions.}

\emph{Proof:} Regarding the objective function, the second-order derivative of the first term $w_nE_{n,s}^t$ w.r.t. $\tau_{n,s}$ is always greater than zero, indicating the convexity for $\tau_{n,s}$. The second and third terms are convex functions w.r.t. $f_m^n$ and $f_m^s$, respectively. Since the objective function is the sum of convex functions, it is also convex. As for the constraints, the function $\frac{C_m^n}{f^{V}_{m}}$ in the first term of (8b) is convex with $f^{V}_{m}$ and constraint (11) is affine function of $f^{V}_{m}$ that is also convex. Additionally, (8b) and (8c) are all affine and convex with $\tau_{n,s}$. This completes the proof.
$\hfill\blacksquare$

The convexity of this problem indicates that we can directly derive globally optimal solution that meets the delay requirements while minimizing vehicle energy consumption. With the satisfaction of Slater's condition, it can be efficiently solved by the Karush-Kuhn-Tucker (KKT) conditions algorithm, which is well suited for real-time decision-making demand in vehicular networks. Then we can derive the optimal solutions of $\tau_{n,s}$ and $f^{V}_{m}$ as presented in the following.

\emph{Lemma 1:} The optimal V2V transmission delay for each RV-SV pair is:
\begin{equation*}
\tau^*_{n,s}=\frac {W_{m_s^n-1,m_s^n}^n}{B_{V2V} \left [{W_{0} \left ({\frac {\frac{10^{-\frac{\phi_{n,s}}{10}} h_{n,s}\lambda }{w_n N_0B_{V2V}}-1}{e}}\right)+1 }\right]}. \tag{12}
\end{equation*}

\emph{Proof:} The KKT conditions formulas with fixed offloading decisions can be listed as following:
\begin{align*}
&\text{Constraints (8b), (8c) and (11),}  \\
&\lambda \left( \sum_{m=1}^{M^n} \frac{C_m^n}{f^{V}_{m}}+\tau_{n,s} - T_{n,max}\right) = 0, \hspace{1.8cm} \tag{13a}\\
&\eta \tau_{n,s} = 0, \tag{13b}\\
&\mu_m f_m^V = 0, m=1,\ldots,M^n, \tag{13c}
\end{align*}
\begin{align*}
&\nu_m \left(f_m^V-F_V\right) = 0, m=1,\ldots,M^n, \tag{13d}\\
&\frac {w_nB_{V2V}N_0}{10^{-\frac{\phi_{n,s}}{10}} h_{n,s}} \left( e^{\frac {W_{m_s^n-1,m_s^n}^n}{\tau_{n,s}B_{V2V}}}-1\right) \\
&-\frac {w_n N_0 W_{m_s^n-1,m_s^n}^n e^{\frac {W_{m_s^n-1,m_s^n}^n }{\tau_{n,s}B_{V2V}}}}{10^{-\frac{\phi_{n,s}}{10}} h_{n,s}\tau_{n,s}} +\lambda -\eta =0, \tag{13e}\\
&2w_n\kappa_{n} C_m^n f^{V}_{m} -\lambda \frac{C_m^n}{\left ({f^{V}_{m}}\right)^{2}} -\mu _{m} + \nu _{m}=0,\\
&m=1,\ldots, m_s^n-1, \tag{13f}\\
&2w_s\kappa_s C_m^n f^{V}_{m} -\lambda \frac{C_m^n}{\left ({f^{V}_{m}}\right)^{2}} -\mu _{m} + \nu _{m}=0,\\
&m=m_s^n,\ldots, M^n. \tag{13g}
\end{align*}
where $\lambda$, $\eta$, $\mu_m$ and $\nu_m$ are the nonnegative Lagrange multipliers associated with the constraint (8b), (8c), $f_m^V\geq 0$ and $f_m^V\leq F_V$, respectively. We have $\tau_{n,s}>0$ to guarantee the rationality. Thus, we have $\eta=0$ based on (13b). We can define $\beta \triangleq\frac{W_{m_s^n-1,m_s^n}^n}{B_{V2V}\tau_{n,s}}$. Hence, (13e) can be rewritten as
\vspace{-0.2cm}
\begin{equation*}
(\beta - 1)e^{\beta - 1 }=\frac{\frac{10^{-\frac{\phi_{n,s}}{10}} h_{n,s}\lambda}{w_n N_0 B_{V2V}}-1}{e}. \tag{14}
\end{equation*}
In order to obtain the optimal expression of $\tau_{n,s}$, we leverage the function $W_0(x)$ whose definition is given in TABLE~\ref{tab:notations}. Since $\beta > 0$, we have $\beta - 1 > -1$. Therefore, we can obtain
\begin{equation*}
\beta = W_0\left(\frac{\frac{10^{-\frac{\phi_{n,s}}{10}} h_{n,s}\lambda}{w_n N_0 B_{V2V}}-1}{e}\right)+1. \tag{15}
\end{equation*}
Then we substitute the defined expression of $\beta$ into (15) and we can finally get the optimal solution as given in (12).$\hfill\blacksquare$

\emph{Lemma 2:} The optimal allocation of computing resource for each subtask by RV and SV is:
\begin{equation*}
{f_m^V}^*=\min \left ({\left ({\frac {\lambda }{2w_{g(m)}\kappa _{g(m)}}}\right)^{\frac {1}{3}}, F_V}\right), m=1,\ldots, M^n, \tag{16}
\end{equation*}
where $g(m) = n, 1 \leq m \leq m_s^n-1$ and $g(m) = s, m_s^n \leq m \leq M^n$. The optimal value of $\lambda$ for solution (12) and (16) can be easily found using the bisection searching method.

\emph{Proof:} We first derive the optimal solution for $\left\{f^{V}_{m}|m=1,\ldots,m_s^n-1 \right\}$. Since each subtask must be computed, we have $f_m^V > 0$. Based on (13c), $\mu_m=0$. When the computing resource constraint (11) is inactive (i.e., $f^{V}_{m}<F_V$), $\nu_m = 0$ from (13d). According to KKT condition (13f), the unconstrained optimal solution is derived as
\begin{equation*}
f_m^V=\left(\frac{\lambda}{2w_n\kappa_n}\right)^{\frac{1}{3}}. \tag{17}
\end{equation*}
Given that $\lambda$, $w_n$, and $\kappa_n$ are unique for each RV–SV pair, all the subtasks $m \in \{1,\ldots,m_s^n-1\}$ follow the same expression. When constraint (11) is active, then $f^{V}_{m} = F_V$. Due to the convexity of the objective function and linear boundary of the feasible set, the optimal computing resource allocation is the projection of the unconstrained stationary point (17) onto the feasible interval $[0, F_V]$. Following the similar method, the solution for $\left\{f^{V}_{m}|m=m_s^n,\ldots,M^n \right\}$ can also be derived. Thus, the unified optimal solution in (16) is obtained.

By this point, we express the optimal V2V transmission delay and computing resource allocation policy as the functions of $\lambda$. Then we analyze and describe the method for deriving the optimal value of $\lambda$. In the objective function of $\mathcal{P}_1$, the first term is monotonically decreasing with $\tau_{n,s}$ when $\tau_{n,s}>0$ (this can be easily proved by second-order derivative) and the rest two terms are monotonically increasing with $f_m^n$ and $f_m^s$, respectively. Based on (12), $\tau_{n,s}^*$ is monotonically decreasing with $\lambda$, while (16) indicating that ${f_m^V}^*$ is monotonically non-decreasing with $\lambda$. Therefore, the objective function is monotonically increasing with $\lambda$. To minimize the objective value, $\lambda$ should be as small as possible. As for the constraint (8b), the function in the left hand side is monotonically decreasing with $\lambda$. Thus, we only need to find the $\lambda$ that just activates the inequality (8b). Due to the monotonicity, this value can be easily found using the bisection searching method. This completes the proof. $\hfill\blacksquare$


\subsection{Complexity Analysis for the Vehicle Tier}

In the RV-SV matching algorithm, the candidate matching identification needs to traverse all RV-IV pairs to verify whether both conditions are satisfied, with the complexity of $\mathcal{O}\left( N I \right)$. The worst-case complexity of the maximum matching step is $\mathcal{O}\left( N^2I \right)$ when all the RV-IV pairs are identified as candidate matching. Hence, the overall complexity of the RV-SV matching algorithm is $\mathcal{O}\left( N^2I \right)$. Obtaining the optimal task offloading decision for matched RVs requires traversing their subtasks, resulting in a complexity of $\mathcal{O}\left( \overline{M}S \right)$, where $\overline{M}$ represents the average number of subtasks for each application, and the number of SVs $S$ is equal to the number of RVs successfully matched in the Vehicle Tier. Due to efficient closed-form solutions, the complexity of deriving the optimal V2V transmission delay and computing resource allocation stems only from the bisection search with tolerance $\epsilon$, yielding a complexity of $\mathcal{O}\left( \log_2(\frac{1}{\epsilon}) \right)$. Consequently, the complexity of optimizing offloading decisions and resource allocation is $\mathcal{O}\left( \overline{M}S \log_2(\frac{1}{\epsilon}) \right)$. Since $\overline{M}$ and $\log(\frac{1}{\epsilon})$ are typically small and $S \leq N$, the dominant complexity of the overall solution for the Vehicle Tier is $\mathcal{O}(N^2 I)$. This indicates that our method scales polynomially with the number of RVs and is computationally efficient for task and resource scheduling in vehicular networks.


\section{Tier-2: Collaborative Computation in the RSU Tier}

Building upon the collaborative computation in the Vehicle Tier, this section investigates the RSU Tier to serve the remaining RVs. Specifically, RVs that fail to be matched with a qualified SVs will offload their application tasks to RSUs. To ensure energy-efficient and timely task completion, we explore the assignment of each RV's subtasks among RSUs, along with the multi-access allocation of RSU computing resources and V2I subchannels.

\subsection{Problem Formulation of the RSU Tier}

In this subsection, we first formulate the problem of the RSU Tier and then provide a detailed problem description. We symbolize the set of RVs that fail to be matched in the Vehicle Tier by $\mathcal{N}'$ with a total number of $N'$. In this work, we consider that each RSU can process tasks of multiple RVs in parallel and optimize the multi-access allocation of RSU computing resources with the assumption that tasks can begin computation immediately upon arrival. We formulate the problem $\mathcal{P}_2$ of the RSU Tier as

$\mathcal{P}_2$: \textbf{Collaborative Computation of RSUs}
\begin{align*}
\min _{\substack { \mathbf{m^n_r} , \mathbf{\tau_n},\\ \mathbf{f_{n,m}^r}, \mathbf{b_n}}} \hspace{0.3cm} &\sum^{N'}_{n=1} \left[ w_n E_n^t + \sum_{r=1}^{R} w_r \left( E^t_{r,n} + \sum_{m=m_r^n}^{m_{r+1}^n-1} E_{n,m}^c \right) \right]\\
\text {s.t. } \hspace{0.4cm} & 1 \leq m_r^n \leq m_{r+1}^n \leq M^n +1, \forall n \in \mathcal {N'}, \\
& r \in \mathcal{R},  \tag{18a}\\
& \sum_{q=1}^{Q^n} \frac{D^n_q}{R_{n}} + \tau_n+\sum_{r=1}^{R}\sum_{m = m^n_r}^{m^n_{r+1}-1}\tau^{c}_{n,m} + \sum_{r=1}^{R-1}\tau_{r,n} \\
&\leq T_{n,max}, \forall n \in \mathcal{N'}, \tag{18b}\\
& \sum_{q=1}^{Q^n} \frac{D^n_q}{R_{n}} + \tau_n \leq \frac{L-S_n}{v_n}, \forall n \in \mathcal{N'}, \tag{18c}\\
& \tau_n \geq 0, \forall n \in \mathcal{N'}, \tag{18d}\\
& \sum_{n=1}^{N'} b_n \leq b, b_n \in \mathcal{I}_+, \tag{18e}\\
& \sum_{n=1}^{N'} f_{n,m}^r \leq F_r, \forall r \in \mathcal{R}. \tag{18f}
\end{align*}

The objective is minimizing the total energy consumption for all RVs in set $\mathcal{N'}$, which encompasses V2I data transmission energy from RVs to the current RSU, inter-RSU wired transmission energy, and RSU computation energy. $w_n$ and $w_r$ are the weights of RV$_n$ and RSU$_r$, respectively. This problem involves four sets of optimization variables: task offloading decisions $\mathbf{m_r^n} = \left\{ m^n_r|n \in \mathcal{N'}, r \in \mathcal{R} \right\}$, V2I transmission delays $\mathbf{\tau_n} = \left\{ \tau_n|n \in \mathcal{N'} \right\}$, RSU computing resource allocation $\mathbf{f_{n,m}^r} = \left\{ f_{n,m}^r | m \in \mathcal{M}^n,n \in \mathcal{N'},r \in \mathcal{R} \right\}$ for each subtask, and the number of subchannels $\mathbf{b_n} = \left \{b_n|n \in \mathcal {N'}\right \}$ allocated to each V2I link.

As for the constraints, (18a) regulates the range and sequence of offloading decisions. Note that $m^n_r=m^n_{r+1}$ means the RSU$_r$ is not assigned any subtask of RV$_n$ and only needs to transmit its intermediate data to RSU$_{r+1}$. When $m^n_r=M^n+1$, it means previous RSUs have completed all the subtasks of RV$_n$. (18b) guarantees the total delay of each RV$_n$'s application task will not exceed its delay requirement, where the first term is the communication setup delay between RV$_n$ and the current RSU. $Q^n$ signifies the total number of communication setup messages of RV$_n$, $D^n_q$ denotes the data size of the $q\mbox{-}th$ setup message, and $R_{n}$ is the data rate of the link dedicated for RV$_n$'s setup message transmission. (18c) ensures that communication setup and data uploading are completed before each RV exits the service range $L$ of the current RSU, where $S_n$ denotes the distance from the starting point of service range to the current position of RV$_n$. (18d) ensures the transmission delays are nonnegative. (18e) enforces the total number of subchannels allocated to V2I links should not exceed the available subchannel number $b$ of the current RSU, where $\mathcal{I}_+$ denotes the set of positive integers. (18f) ensures the total computing resources allocated to RVs are bounded by the available capacity of each RSU.

The problem $\mathcal{P}_2$ is highly intricate, coupling two types of discrete variables with two types of continuous variables. To tackle this, we propose a layered optimization framework. In the lower layer, with fixed task offloading decisions for RVs, the convex optimization techniques are employed to derive the optimal continuous solutions for V2I transmission delay, V2I bandwidth allocation, and RSU computing resource allocation. In the upper layer, the task offloading decisions can be efficiently determined through linear search. The variables in the lower and upper layers are iteratively updated until they converge to a stable strategy. Subsequently, a gain priority-based scheme is designed to recover the discrete V2I subchannel allocation.


\subsection{Continuous Solutions to Communication and Computing Resource Allocation}

In this subsection, we elaborate on the approach for obtaining the optimal continuous solutions for V2I transmission delay, bandwidth allocation for V2I links, and RSU computing resource allocation.

To enhance the practicality of our scheme, V2I bandwidth is allocated via discrete subchannels. However, directly optimizing the integer set $\mathbf{b_n}$ is an NP-Hard problem. Thus, we propose a two-step strategy, where we first relax the discrete term $b_n B_0$ into a continuous variable $B_n$ to obtain the optimal continuous bandwidth allocation. Subsequently, the near-optimal integer solution $\mathbf{b_n}$ is recovered through a gain priority-based method. Consequently, constraint (18e) is transformed into:
\begin{align*}
& \sum_{n=1}^{N'} B_n \leq B, \tag{19a} \\
& B_n \geq 0, n \in \mathcal{N'}. \tag{19b} 
\end{align*}

\textbf{Proposition 2:} \textit{After relaxing the discrete variables of subchannel allocation, $\mathcal{P}_2$ is a convex optimization problem with fixed task offloading decisions.}

\emph{Proof:} For the objective function, its Hessian matrix w.r.t. $\tau_n$ and $B_n$ is positive-definite, proving its convexity. The second term is determined with given offloading decisions while the third term is convex with $f_{n,m}^r$. Thus, the objective function is convex. Regarding the constraints, function $\frac{C^n_m}{f_{n,m}^r}$ in (18b) is convex with $f_{n,m}^r$ and (18f) is affine function of $f_{n,m}^r$ that is also convex. (18b), (18c), and (18d) are all affine and convex with $\tau_n$. (19) are all affine and convex with $B_n$. This completes the proof. $\hfill\blacksquare$

\subsubsection{Optimal Communication Resource Allocation} We can derive the optimal communication resource allocation strategies as presented in the following.

\emph{Lemma 3:} The optimal solutions of the V2I transmission delay, V2I bandwidth allocation, and computing resources allocation policy are given by
\begin{equation*}
\tau^*_{n}= \min \Bigg\{\frac {W^{n}_{0}}{B_{n} \left [ W_{0} \left (  \left( \frac{\lambda_n \phi_n h_n}{w_n N_0 B_{n}}-1 \right) e^{-1} \right)+1 \right]}, \tau_n^{re} \Bigg\}, \tag{20a}
\end{equation*}
\begin{equation*}
B^*_{n}=\frac {W^{n}_{0}}{\tau_n \left [{W_{0} \left( \left ( \frac{\xi \phi_n h_n}{w_n N_0 \tau_n}-1 \right) e^{-1}\right)+1 }\right]}, \tag{20b}
\end{equation*}
where $\tau_n^{re} = (L-S_n)/v_n - \sum_{q=1}^{Q^n} D^n_q/R_{n}$ is the remaining dwell time of RV$_n$ within the coverage of the current RSU, $\lambda_n$ and $\xi$ are nonnegative Lagrange multipliers associated with the constraint (18b) and (19a), respectively.

\emph{Proof:} The derivation is similar to that of (12) and is omitted here for brevity.
$\hfill\blacksquare$

\subsubsection{Optimal Computing Resource Allocation}

As for the computing resource allocation, the optimal $f_{n,m}^r$ is achieved when the following KKT conditions hold:
\begin{equation*}
\varphi_r \left( \sum_{n \in \mathcal{N}'} f_{n,m}^r - F_r \right) = 0, \tag{21a}
\end{equation*}
\begin{equation*}
2 w_r \kappa_r C_m^n f_{n,m}^r -\lambda_n \frac{C_m^n}{\left ( f_{n,m}^r \right)^{2}} + \varphi_r = 0, \tag{21b}
\end{equation*}
where $\varphi_r$ is the Lagrange multiplier associated with constraint (18f) of each RSU. When (18f) is inactive, from (21a), $\varphi_r = 0$. Then the optimal solution for $f_{n,m}^r$ is given by
\begin{equation*}
{f_{n,m}^r}^*=\left ({\frac {\lambda_n}{2w_r\kappa _{r}}}\right)^{\frac {1}{3}}.  \tag{22}
\end{equation*}
When (18f) is active (i.e., the total computing resources allocated by RSU$_r$ following (22) exceed its capacity), the optimal ${f_{n,m}^r}^*$ is the non-negative root of (21b). We can rewrite (21b) as $2 w_r \kappa_r C_m^n (f_{n,m}^r)^3 + \varphi_r (f_{n,m}^r)^2 -\lambda_n C_m^n = 0$. Since its LHS is monotonically increasing when $f_{n,m}^r > 0$ and negative at $f_{n,m}^r = 0$, it has a unique positive real root. As $f_{n,m}^r$ decreases with $\varphi_r$, the optimal $\varphi_r$ that ensures (18f) is just satisfied can be easily obtained via bisection search.

\subsubsection{Derivation of Optimal Lagrange Multipliers}

Up to this point, we have derived the optimal continuous solutions for communication and computing resource allocation, parameterized by multiple types of Lagrange multipliers. Next, we present the approach to obtain the optimal values of these Lagrange multipliers.

\begin{enumerate}

\item \textit{Initialization of $\lambda_n$:} At the beginning, an initial value of $\lambda_n$ is assigned to each RV$_n$.

\item \textit{Update of Communication Resource Allocation:} The optimal $B_n^*$ is achieved only when the total bandwidth $B$ is fully utilized, activating constraint (19a). Otherwise, allocating any residual bandwidth would further reduce V2I transmission energy. Consequently, we should find a $\xi$ satisfying $\sum_{n=1}^{N'} B_n = B$. Since the LHS of (19a) monotonically decreases with $\xi$, the optimal $\xi$ can be easily determined via bisection search. Given the values of $\lambda_n$ and $\xi$, the coupled variables $\tau_n$ and $B_n$ can be updated alternately until convergence to stable values.

\item \textit{Update of Computing Resource Allocation:} The computing resources allocation of each RSU is obtained according to (22). Then constraints (18f) for all RSUs are checked. If all of them are satisfied, we proceed to the next step. Otherwise, if the total allocated computing resources at any RSU exceed its capacity, a bisection search is employed to determine the value of $\varphi_r$ that just activates (18f), thereby obtaining its computing resource allocation.

\item \textit{Update of $\lambda_n$:} According to the analysis in Section III-C, the value of $\lambda_n$ for each RV$_n$ is updated such that (18b) is just active. Steps 2 – 4 are then repeated until the overall resource allocation strategy converges.
    
\end{enumerate}

With given $\lambda_n$, the communication and computing resource allocation are decoupled and can be solved in parallel. Owing to the convexity, the above procedure will ultimately converges to a stable optimal overall solution.


\subsection{Solutions to Task Offloading Decisions and Subchannel Allocation}

Thus far, we have obtained the optimal continuous solutions for V2I transmission delay, V2I bandwidth allocation, and RSU computing resource allocation in the lower layer of the optimization framework. Building upon this, the sequential subtask offloading decisions for RVs in the upper layer can be determined through linear search method.

As for the subchannel allocation of V2I links, instead of directly tackling the NP-hard discrete optimization problem in $\mathcal{P}_2$, we design a low-complexity method to recover the discrete solution based on the optimal continuous bandwidth allocation $B_n^*$. The optimal continuous number of subchannels for each V2I link is given by $b_n^* = B_n^* / B_0$. Since the V2I transmission energy $E_n^t$ in objective function is a separable convex function of the subchannel number $b_n$, the optimal solution in discrete domain naturally lies in the integers adjacent to each continuous $b_n^*$. Motivated by this, a gain priority-based subchannel allocation strategy is designed, with the procedure described as follows:
\begin{enumerate}
    \item \textit{Initial Quantization:} For each RV$_n$, the optimal continuous solution $b_n^*$ is floored to obtain the initial discrete solution $\hat{b}_n = \lfloor b_n^* \rfloor$.
    
    \item \textit{Gap Calculation:} The number of remaining subchannels $k$ resulting from the flooring operation is calculated by $k = b - \sum_{n=1}^{N'} \hat{b}_n$. Since $0 \leq b_n^* - \hat{b}_n < 1$, we have $0 \le k < N'$.
    
    \item \textit{Gain Priority-Based Subchannel Allocation:} The gain priority of each RV$_n$ is calculated by $G_n = w_n \left( E_n^t(\hat{b}_n) - E_n^t(\hat{b}_n + 1) \right)$, which is defined as the reduction in V2I transmission energy if it is allocated one additional subchannel. To minimize the total V2I transmission energy, the top $k$ RVs ranked by $G_n$ are allocated $\hat{b}_n + 1$ subchannels, while the remaining RVs are allocated $\hat{b}_n$ subchannels.
    
\end{enumerate}
Through the above process, we conserve the V2I transmission energy to the greatest extent with low complexity while ensuring that the total allocated subchannels do not exceed the current RSU's capacity.


\subsection{Overall Optimization Framework}

\begin{figure}[!t]
\centering
\includegraphics[width=3.4 in]{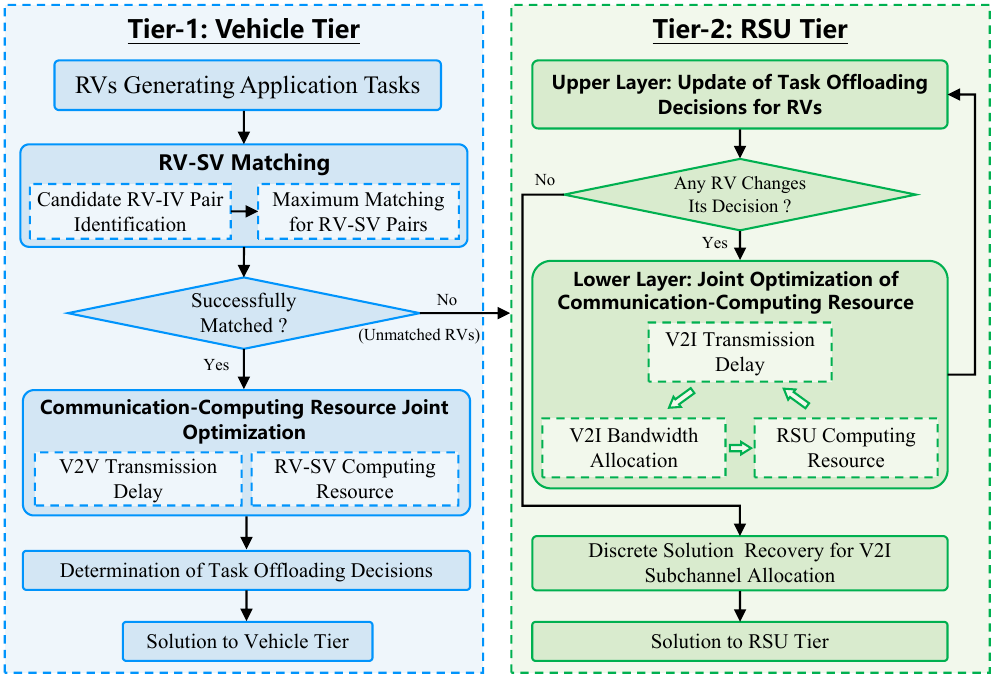}
\caption{Illustration of the overall optimization framework for the proposed method.}
\label{overall_solution}
\end{figure}

The overall optimization procedure of the RSU Tier is summarized as follows: At the initial stage, each RV determines its subtask assignment associated with minimum total energy cost while satisfying task latency constraint via linear search, assuming each RSU allocates all computing resources to each RV. Then the initial offloading decision $\mathbf{m_r^{n(0)}} = \{ \mathbf{m_r^{1(0)}}, ..., \mathbf{m_r^{N'(0)}} \}$ is obtained. Based on the current subtask assignment, the V2I transmission delay, bandwidth allocation, and RSU computing resource allocation are jointly optimized using the method detailed in Section IV-B. Subsequently, RVs update their task offloading decisions according to the optimized resource allocation strategy. Thereafter, the resource allocation optimization and offloading decision update are iteratively performed until no RV changes its decision. Finally, the V2I subchannel allocation scheme is executed, thereby yielding the ultimate solution. To provide a clear presentation, the overall optimization procedure for the Vehicle Tier and the RSU Tier is illustrated in Fig.~\ref{overall_solution}.


\subsection{Complexity Analysis for the RSU Tier}
In this subsection, we conduct the complexity analysis of the proposed method under practical scenarios. Typically, the service range of an RSU is around 200 m \cite{zxiao2023perception}. A vehicle driving at tens of km/h takes roughly 10 seconds to traverse the coverage, which far exceeds the delay requirement (i.e., hundreds of milliseconds) of targeted tasks. This implies that an RV can enter at most two RSUs' service ranges during task execution. To fully exploit computing resources of nearby RSUs, it is reasonable to consider three RSUs' collaboration, while our method remains scalable to more RSUs. In the lower layer of the optimization framework, the communication resource allocation is updated via a bisection search for the optimal $\xi$  with tolerance $\epsilon$, resulting in a complexity of $\mathcal{O}\left( \log_2(\frac{1}{\epsilon}) \right)$. For the computing resource allocation, a bisection search for optimal $\varphi_r$ is required for all RSUs in the worst case, with a complexity of $\mathcal{O}\left( 3 \log_2(\frac{1}{\epsilon}) \right)$. Obtaining optimal $\lambda_n$ of all RVs incurs the complexity of $\mathcal{O}\left( N' \log_2(\frac{1}{\epsilon}) \right)$. Thus, the dominant complexity of the above process is $\mathcal{O}\left( I_1 N' \log_2(\frac{1}{\epsilon}) \right)$, where $I_1$ is the iteration number of lower layer. In the upper layer, the complexity of determining task offloading decisions in the worst case is $\mathcal{O}\left( \overline{M}^2 N' \right)$, where $\overline{M}$ denotes the average number of subtasks per application. The subchannel allocation mainly involves sorting RV gains, with a complexity of $\mathcal{O}\left( N' \log_2 N' \right)$. Hence, the complexity of the overall solution for $\mathcal{P}_2$ is $\mathcal{O}\left( I_2 N' \left( \overline{M}^2 + I_1 \log_2(\frac{1}{\epsilon}) \right) + N' \log_2 N' \right)$, where $I_2$ is the iteration number of upper layer. Since $I_1$ and $\log_2(\frac{1}{\epsilon})$ are typically small, the dominant complexity is $\mathcal{O}\left( I_2 N' \overline{M}^2 + N' \log_2 N' \right)$. This near-linear complexity demonstrates excellent scalability of our approach, enabling timely resource scheduling under massive connectivity in vehicular networks.


\section{Simulations}
In this section, we conduct extensive experiments to evaluate the performance of the proposed method through the comparison with benchmark methods and examine the impact of various system parameters on its performance.

Regarding the application with sequential subtasks, we consider the image classification application\cite{akrizhevsky2012imagenet} with eight-layer DNN, where the concatenated layers can be regarded as sequential subtasks. The input data size and computational workload of each subtask correspond to the amount of data processed and CPU cycles required by each layer, respectively. These can be estimated from parameters such as feature map dimensions, sizes of convolution kernels and pooling windows, etc. To evaluate the robustness of the proposed method under diverse tasks, similar to \cite{qshen2022dependency}, the input data size and computational workload of each subtask are uniformly distributed within $[1, 20]$ Mb and $[1, 1000]\times10^6$ CPU cycles, respectively. The noise power spectral density is set to $N_0=-140$ dBW/Hz. The default delay requirement for each application is initialized as $0.2$ s near the average human reaction time. 
\vspace{-0.3cm}


\subsection{Experiment for the Vehicle Tier}

In this subsection, we present the experimental settings and results of the Vehicle Tier. We consider a three-lane unidirectional road with lane width of $3.75$ m \cite{zxiao2023perception}. The vehicle positions follow the Poisson point distribution\cite{hguo2022v2v}. Vehicle speeds span from $40$ to $80$ km/h and V2V communication range is set to $70$ m \cite{zxiao2023perception}. The maximum computing resources and computation energy efficiency coefficients of vehicles are configured within $[1,10]$ GHz and $ [1, 2]\times 10^{-23}$, respectively \cite{hzhang2024partial}. To better reflect performance differences, the average energy consumption (AEC) of each task is normalized by the default V2V bandwidth (i.e., $10$ MHz).

\begin{figure}[!t]
\centering
\includegraphics[width=2.8 in]{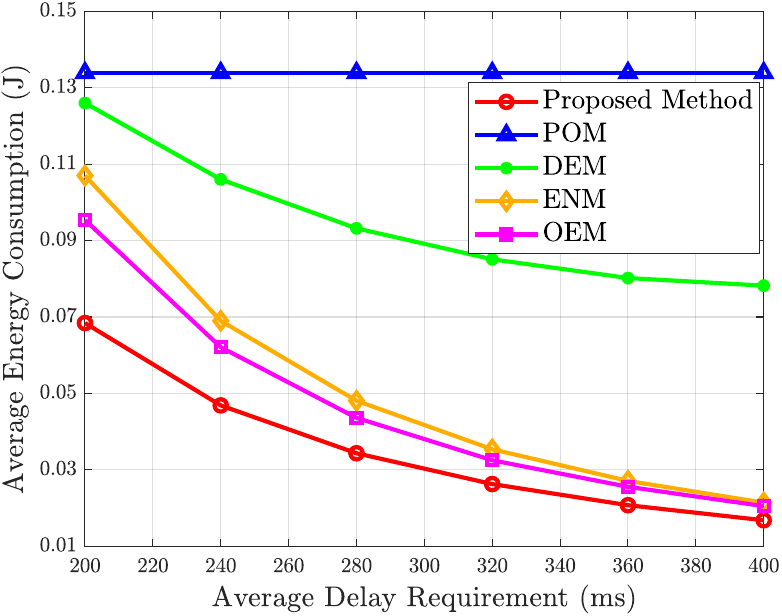}
\caption{Average energy consumption for different methods under varying task delay requirement.}
\label{vt_benchmark}
\end{figure}

\textbf{(1) AEC for different methods under varying task delay requirements:} Fig.~\ref{vt_benchmark} illustrates the AEC among different methods under varying tolerable task delay. We compare the performance of the proposed method with the following benchmark schemes:

\begin{itemize}

\item \textbf{Partial Offloading $(50\%)$ with Maximal CPU Frequency (POM)}: Each pair of RV and SV undertakes half of the subtasks and the computation is performed at their maximum CPU frequency.

\item \textbf{Delay Minimization Scheme (DEM)}: The task execution delay minimization algorithm proposed by \cite{yzhang2025latency} within a vehicle-RSU collaborative architecture.

\item \textbf{Energy Minimization Scheme (ENM)}: The task energy consumption minimization method of \cite{xwang2024amtos} with full priority assigned to energy saving.

\item \textbf{Offloading Efficiency Maximization Scheme (OEM)}: The dependency-aware task offloading and resource allocation scheme that aims to maximize offloading efficiency proposed in \cite{qshen2022dependency}.

\end{itemize}

The experimental results validate that the proposed method consistently achieves the minimum AEC across varying task delay requirements, demonstrating its effectiveness and robustness under diverse task urgencies. Moreover, the advantages of the proposed method are increasingly significant as the delay requirements become more stringent. POM guides the vehicles utilize their full computing capability to process tasks, which is highly energy-consuming and results in the highest AEC. With fixed task workloads, operating at the maximum CPU frequency leads to a dominant and constant computation energy consumption, making the energy cost of POM relatively insensitive to the changes in task delay requirement. DEM fully utilizes the transmit power of vehicles and system computing resources to pursue minimum task execution delay, which leads to considerable energy consumption for both data transfer and computation. Although ENM aims at minimizing energy cost, treating each application task as an indivisible whole limits scheduling flexibility. The workload imbalance causes vehicles assigned heavier workloads to consume high computation energy. By allocating subtasks to two types of vehicles in a finer-grained manner, the proposed method achieves around 29\% reduction in AEC compared with ENM. Due to multi-hop task offloading, OEM incurs extra communication latency and energy, and further consumes more computing resources and energy to meet delay constraints, enabling the proposed method reduces the AEC by around 23\% compared with it.

\begin{figure}[!t]
\centering
\includegraphics[width=2.7 in]{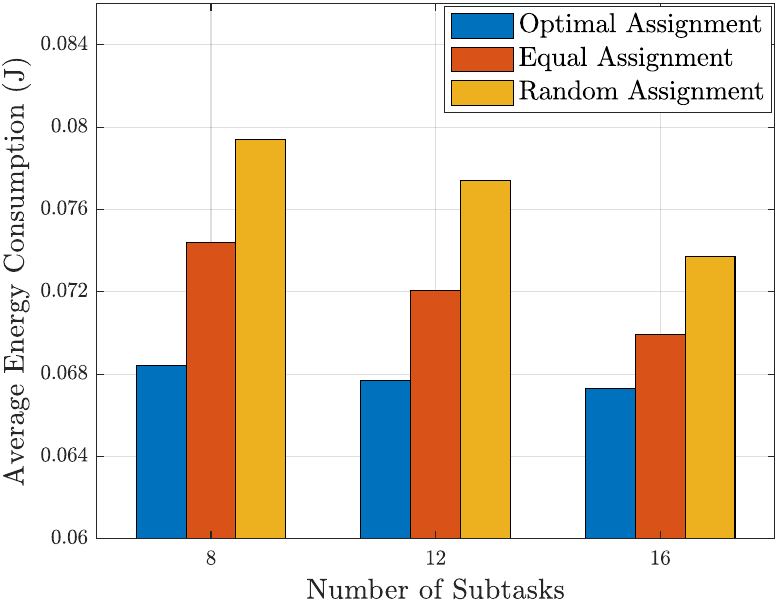}
\caption{Average energy consumption with varying subtask structure.}
\label{vt_subtask_structure}
\end{figure}

\textbf{(2) Impact of subtask structure:} In Fig.~\ref{vt_subtask_structure}, we fix the total data size and computational workloads of the application tasks and change the number of subtasks by further splitting some of them. For each subtask number, different offloading decisions (i.e., optimal, equal, and random assignment) are considered. The optimal assignment consistently achieves the lowest AEC, whereas random assignment leads to significantly higher AEC due to its inherent uncertainty. Furthermore, a general trend can be observed that finer-grained subtask partitioning provides the proposed method with greater scheduling flexibility, thereby achieving lower AEC. This conclusion is supported by extensive experiments and holds for most scenarios, although exceptions may exist in certain special cases.

\begin{figure}[!t]
\centering
\includegraphics[width=2.8 in]{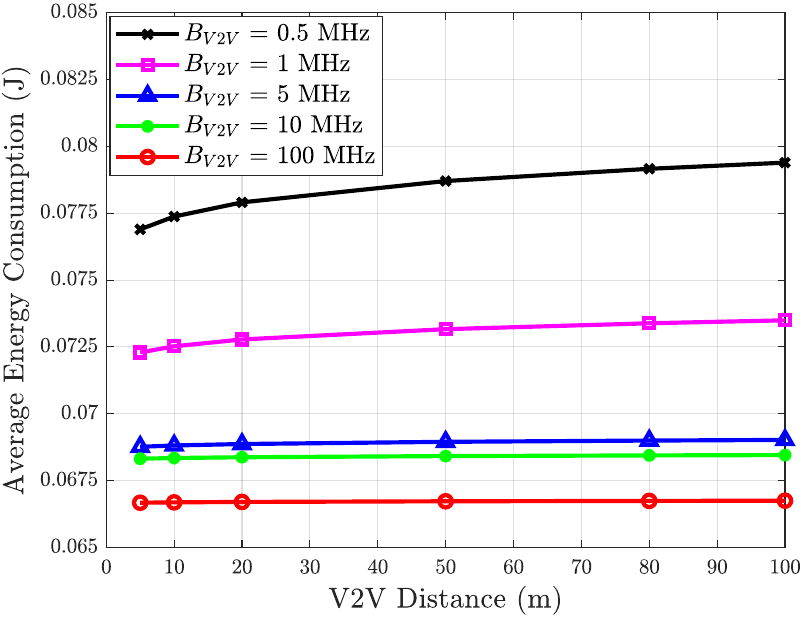}
\caption{Impact of V2V bandwidth and distance on average energy consumption.}
\label{vt_v2v_bandwidth}
\end{figure}

\textbf{(3) Impact of V2V bandwidth and distance between RV and SV:} In the scenario of Fig.~\ref{vt_v2v_bandwidth}, $B_{V2V}$ is set to $0.5$, $1$, $5$, $10$, and $100$ MHz. The AEC decreases continuously with increasing $B_{V2V}$, which can be interpreted from two perspectives: On the one hand, a larger $B_{V2V}$ requires lower transmit power to achieve the same data rate. On the other hand, under fixed transmit power, a larger bandwidth increases the data rate and reduces the transmission delay. Consequently, more time is available for task computation, allowing vehicles to operate at lower CPU frequencies with reduced energy cost. In addition, for a fixed $B_{V2V}$, larger inter-vehicle distances result in higher AEC due to the increased transmit power requirement. However, this increase remains limited since transmission energy accounts for a small portion of the total energy consumption.

\begin{figure}[!t]
\centering
\includegraphics[width=2.8 in]{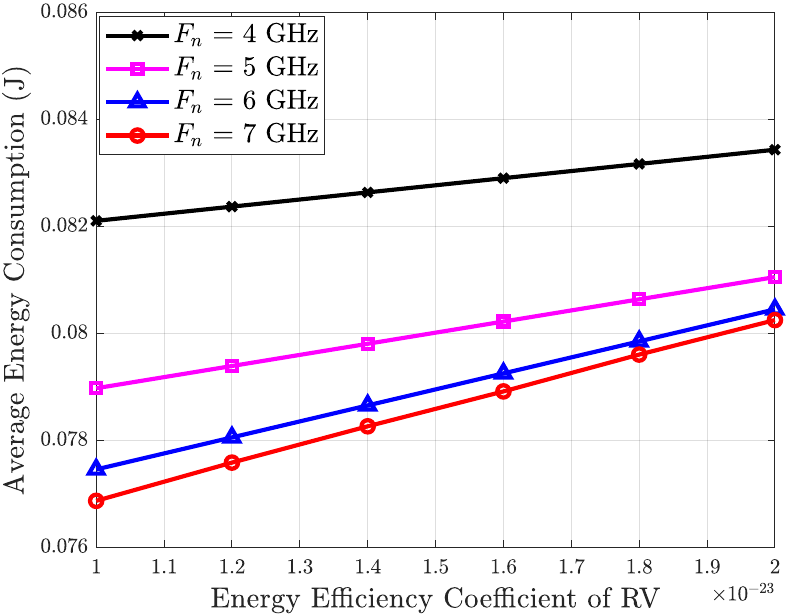}
\caption{Impact of computing capacity and energy efficiency coefficient of RV on average energy consumption.}
\label{vt_rv_computing}
\end{figure}

\textbf{(4) Impact of computing capacity and energy efficiency coefficient of RV:} In Fig.~\ref{vt_rv_computing}, the maximal computing capacity $F_n$ of RV varies from $4$, $5$, $6$, to $7$ GHz, respectively, while that of SV is fixed at $10$ GHz. The AEC decreases as the computing capacity of RV approaches that of SV. This is because a faster computation on RV alleviates the burden of SV, allowing SV to operate at a lower CPU frequency. The increase in RV's computation energy is smaller than the decrease in SV, leading to less total energy consumption. Moreover, a higher computing capacity of an RV also makes its computation energy consumption more sensitive to the energy efficiency coefficient $\kappa$, since it will undertake heavier workloads accordingly.


\subsection{Experiment for the RSU Tier}

This subsection elaborates on the experimental results of the RSU Tier. The RSU service range is set to $200$ m according to \cite{zxiao2023perception}. Similarly to Tier-1, the energy efficiency coefficients of RSUs are configured within $[1, 2]\times 10^{-23}$, while their maximal computing capacities range from $60$ to $120$ GHz. For communication-related parameters, we set the communication setup delay to $0.1$ ms, $E_0=10^{-5}$ J and $\tau_0=10^{-5}$ ms. These parameters can be flexibly adjusted in practical scenarios. The AEC is normalized by the default system bandwidth of $100$ MHz. To fully exploit the computing resources of nearby RSUs, we deploy three RSUs for collaborative computation in this experiment. This method is also scalable to the practical scenarios with a larger number of RSUs.

\begin{figure}[!t]
\centering
\includegraphics[width=2.8 in]{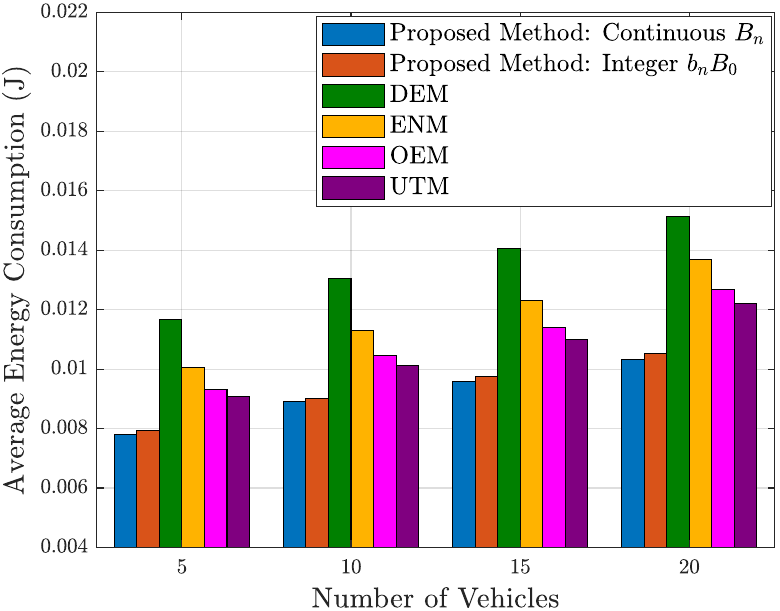}
\caption{Average energy consumption among different methods with varying number of vehicles.}
\label{rt_benchmark}
\end{figure}

\textbf{(1) AEC for different methods with diverse number of RV:} Fig.~\ref{rt_benchmark} shows the AEC of following methods under varying number of RVs. 

\begin{itemize}

\item \textbf{Proposed Method: Continuous $B_n$}: The proposed method where bandwidth $B_n$ is modeled as a continuous variable.

\item \textbf{Proposed Method: Integer $b_n B_0$}: The proposed method implementing discrete subchannel allocation $b_n$ with subchannel bandwidth of $1$ MHz.

\item \textbf{Utility Maximization Scheme (UTM)}: The game theory-based algorithm that maximizes the utility of vehicle users and RSU servers proposed in \cite{lzhao2024stackelberg}.

\item The \textbf{DEM}, \textbf{ENM}, and \textbf{OEM} schemes introduced in the previous subsection.

\end{itemize}

The proposed method with continuous $B_n$ consistently demonstrates the best performance under varying vehicle number as it achieves optimality for all optimization variables. Discretizing $B_n$, as in the second method, slightly increases the AEC due to subtle deviation from the optimal continuous solution. The slight performance degradation compared with the optimal strategy validates the effectiveness of the proposed subchannel allocation scheme. Due to the unoptimized V2I bandwidth allocation, transmission delays, and scattered subtask offloading, UTM and OEM exhibit noticeably higher AEC than our methods. UTM places greater emphasis on energy saving in utility function, thus achieving lower AEC than OEM. Even so, the proposed method still reduces the AEC by approximately 15\% compared with UTM. To minimize task execution latency, DEM fully exploits and coordinates RSU computing resources to accelerate parallel task processing for multiple RVs. This inherently triggers a surge in task energy consumption due to the quadratic relationship between processor frequency and computation energy. While the ENM scheme explicitly optimizes for energy conservation, allocating an entire application task as an indivisible entity to a single RSU concentrates the computational workloads, necessitating higher processor frequencies that inherently incur increased computation energy. In contrast, our proposed framework distributes sequential subtasks across cooperative wired RSUs, effectively balancing the computational workloads and coordinating resources to enable energy-efficient, demand-aware processing while adhering to task delay boundaries.

\begin{figure}[!t]
\centering
\includegraphics[width=2.8 in]{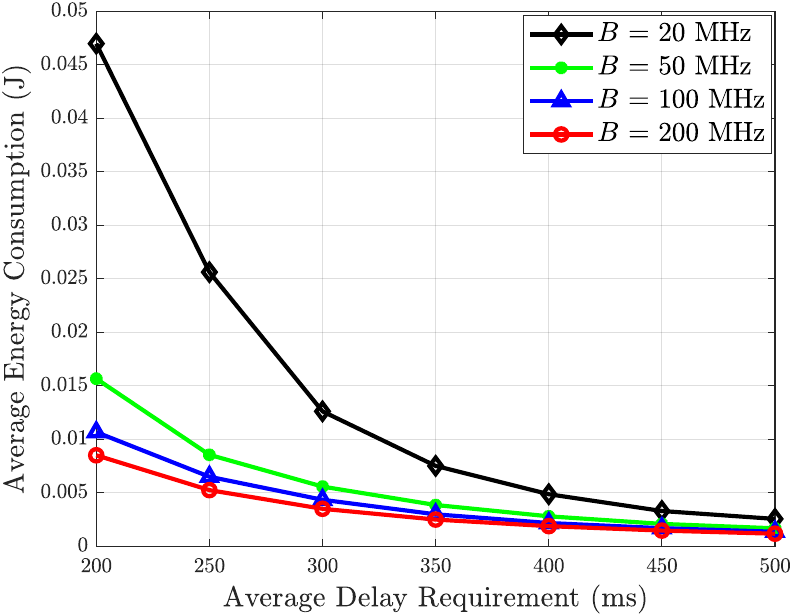}
\caption{Impact of system bandwidth on average energy consumption.}
\label{rt_system_bw}
\end{figure}

\textbf{(2) Impact of system bandwidth:} To evaluate the impact of communication resource availability, the total bandwidth $B$ of the current RSU is configured to $20$, $50$, $100$, and $200$ MHz in Fig.~\ref{rt_system_bw}, respectively. The results demonstrate that expanding the system bandwidth consistently reduces the AEC. This improvement is governed by the fact that a wider bandwidth enhances the V2I uplink transmission capacity, thereby curtailing the communication delay and yielding a larger time margin for the subsequent computation phase. Furthermore, Fig.~\ref{rt_system_bw} illustrates that relaxing the delay requirement significantly decreases the AEC. Inherently, a more generous delay boundary alleviates the demand for the working power in both data transmission and task computation, enabling the system to operate in a more energy-efficient manner.

\begin{figure}[!t]
\centering
\includegraphics[width=2.8 in]{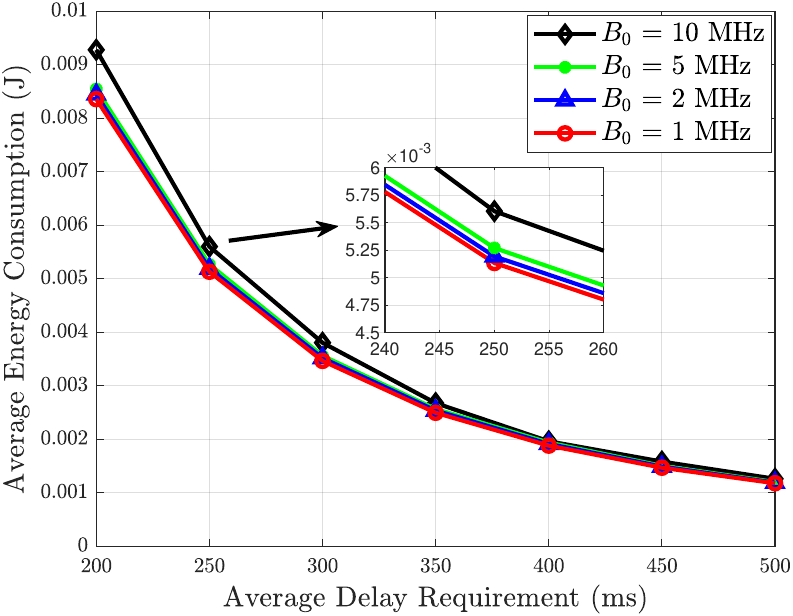}
\caption{Impact of subchannel bandwidth on average energy consumption.}
\label{rt_subchannel_bw}
\end{figure}

\textbf{(3) Impact of subchannel bandwidth:} Fig.~\ref{rt_subchannel_bw} illustrates the impact of subchannel division granularity on the AEC. As the unit subchannel bandwidth $B_0$ increases, the AEC exhibits an upward trend. This performance degradation arises because a coarser partitioning granularity exacerbates the quantization error, causing the discrete bandwidth allocation strategy to deviate further from the optimal continuous solution, thereby compromising overall energy efficiency. Nevertheless, our recovery mechanism for discrete subchannel allocation is executed subsequent to obtaining the optimal continuous solutions, thereby substantially preserving the core optimality of the multidimensional resource joint optimization framework. Furthermore, since data transmission energy accounts for a relatively minor proportion of the total energy consumption, the sensitivity of AEC to bandwidth discretization granularity remains generally acceptable within a reasonable range.


\section{Conclusions}

In this article, we proposed a V2X-empowered multi-tier task offloading mechanism for vehicular applications with sequential subtasks. In the Vehicle Tier, the RV-SV matching scheme effectively accommodates diverse vehicle mobilities and resource heterogeneity to maximize the number of matched RVs. By jointly optimizing offloading decisions alongside communication-computing resource allocation, we minimized vehicle energy cost while strictly ensuring timely task completion. In the RSU Tier, we further investigated the on-demand multi-access challenges of uplink subchannel and RSU computing resources for unmatched RVs, efficiently addressing this complex problem via a layered optimization architecture and a subchannel allocation scheme. Extensive experiments validated the significant energy efficiency superiority of the proposed method over benchmark schemes. These findings provide valuable insights for the development of reliable and sustainable MEC-enabled vehicular networks.


\vfill

\end{document}